# The Case for Repeatable Analysis with Energy Economy Optimization Models

Joseph F. DeCarolis*[1], Kevin Hunter[1], Sarat Sreepathi[2]

[1]*Department of Civil, Construction, and Environmental Engineering*
[2]*Department of Computer Science*
*North Carolina State University*
*Raleigh, NC 27695 United States*

* Corresponding author: Tel.: +1 919 515 0480
Email address: jdecarolis@ncsu.edu

**Abstract**
Energy economy optimization (EEO) models employ formal search techniques to explore the future decision space over several decades in order to deliver policy-relevant insights. EEO models are a critical tool for decision-makers who must make near-term decisions with long-term effects in the face of large future uncertainties. While the number of model-based analyses proliferates, insufficient attention is paid to transparency in model development and application. Given the complex, data-intensive nature of EEO models and the general lack of access to source code and data, many of the assumptions underlying model-based analysis are hidden from external observers. This paper discusses the simplifications and subjective judgments involved in the model building process, which cannot be fully articulated in journal papers, reports, or model documentation. In addition, we argue that for all practical purposes, EEO model-based insights cannot be validated through comparison to real world outcomes. As a result, modelers are left without credible metrics to assess a model's ability to deliver reliable insight. We assert that EEO models should be discoverable through interrogation of publicly available source code and data. In addition, third parties should be able to run a specific model instance in order to independently verify published results. Yet a review of twelve EEO models suggests that in most cases, replication of model results is currently impossible. We provide several recommendations to help develop and sustain a software framework for repeatable model analysis.

## 1 Introduction

Future environmental sustainability – at geographic scales ranging from local to global – is tightly linked to energy system performance. Decision-makers must make difficult choices related to energy and environmental policy as well as energy technology investment in the face of large future uncertainties. While many types of computer-based models are used to perform energy system analysis, this paper focuses on energy economy optimization (EEO) models, which have been used to produce high visibility analyses at the regional, national, and international scale (e.g., Clarke et al., 2007; EIA, 2011a; Nakicenovic et al., 2000; IEA, 2010). Given the concern over anthropogenic climate change, such model-based analyses are being produced at an accelerated pace and are playing an increasingly influential role in the policy-making process.

EEO models employ formal search techniques, which enable the systematic exploration of the future decision space in order to deliver prescriptive, policy-relevant insight and relate near-term actions to long-term outcomes. Although the mathematical formulation varies by model, EEO models optimize consumption and/or energy supply over time in order to minimize the system-wide cost of energy or maximize social utility, subject to constraints representing physical limitations and public policy. While the outputs also vary by model, examples include the optimal portfolio of technologies over time, projected fuel and commodity prices, aggregate economic impact, and emissions of air pollutants and greenhouse gases. In addition, integrated assessment models have extended EEO models to consider the physical impact on the climate system, thereby tying energy system development to anthropogenic climate change (Weyant, 2009).

With nearly all focus placed on model application to generate insight, insufficient attention has been paid to transparency in EEO model development and application. Over the last couple of decades, increasing data availability and dramatic improvements in the cost performance of computer hardware have led to increasingly large and complex EEO models. In addition, long timescales for analysis preclude model validation, so there is little to guide the modeler and reign in efforts that do not improve a model's ability to deliver useful insight (DeCarolis, 2010). As models proliferate and grow more complicated, many modeling teams continue to restrict access to source code and data. A critical consequence is that third parties cannot reproduce published model results. As a result, external reviews of EEO model-based analysis are based largely on comparison to existing literature, domain expertise, and trust in the modeling team performing the work rather than hands-on experience (Ha-Duong, 2001). However, if results are driven by hidden assumptions, coding bugs, highly sensitive parameters, or specific details related to model formulation, they may be impossible to detect by external observers. Given the important role that EEO models play in policy development, we argue strongly that model-based analysis aimed at informing public policy should be repeatable by third parties. Others have echoed the same view, notably the Advisory Group to the European Commission's Energy Roadmap 2050, which notes that the proprietary PRIMES model should be made publicly available since it delivered key results that informed the roadmap (Helm et al., 2011). Furthermore, we assert that repeatable model-based analysis requires publicly accessible code and data, a point that has been

   

persuasively argued in the broader scientific community (Ince et al. 2012; Hanson et al. 2011; Barnes 2010).

With the intention of spurring a discussion of model accessibility standards within the international energy modeling community, this paper is organized into six sections with the following objectives: (1) describe the unique strengths and limitations of EEO models, (2) articulate the need for repeatable analyses with such models, (3) review the availability of some well known models, (4) assess the adequacy of existing open source software licenses, (5) provide a set of recommendations that will enable repeatable experiments with EEO models, and (6) draw conclusions regarding the feasibility of generating repeatable analysis with EEO models.

## 2  EEO models as instruments of inquiry

EEO models – ambitious in scope and data-intensive by nature – combine scientific understanding, codified rules for human behavior, and value judgments into a unified evaluation framework. The expansive system boundaries and complex interactions that determine the trajectory of energy systems require considerable consolidation of real world phenomena into model code and data to enable analysis. The process of distilling relevant literature into a mathematical model requires reasoned judgment that makes modeling as much art as science (Morrison and Morgan, 1999; Ravetz, 2003).

Nordhaus (2008) likens his DICE integrated assessment model to an iceberg, where the "visible portion" is a small set of mathematical equations, which are informed by hundreds of studies that lie "beneath the surface." A prime example is the assumed value for the social discount rate, which has been the subject of much high visibility debate (e.g., Goulder and Stavins, 2002; Nordhaus, 2007; Stern and Taylor, 2007). In a single number, the discount rate can embody the societal weight assigned to current versus future generations, the degree of reversibility of future environmental damages, and the level of optimism regarding future technological development. Generally, complex, dynamic phenomena are radically simplified to make model formulation tractable. For example, technology explicit, partial equilibrium models assume perfect markets for commodities, whereby each commodity is produced at a level that maximizes the sum of producer and consumer surplus (Loulou et al., 2004). In addition, behavioral response and acceptance of new technology is often modeled simplistically as a hurdle rate: a higher, technology-

  3

specific discount rate that represents consumer reluctance to accept newer, less familiar technologies (e.g., Kannan, 2009, Yeh et al., 2006). Likewise, computable general equilibrium models typically assume that the social welfare function is an aggregation of preferences across individuals, without regard to how the consumption is distributed (Ackerman et al., 2009). More broadly, the underlying assumptions employed by economic modelers do not reflect a deep scientific understanding of human behavior, but rather "convenience and convention" (Ackerman et al., 2009).

Despite the simplifying assumptions and subjective judgments required to construct an EEO model, it nonetheless serves as a valuable tool of inquiry that is greater than the sum of its parts. Theory or data in isolation yield little insight about potential future outcomes related to energy technology deployment, energy and environmental policy, and economic growth. However, a dynamic model that applies economic theory in a self-consistent manner to a given dataset can capture interactive effects among technologies and/or between economic sectors to yield new insight from a systems perspective. While EEO models can serve as valuable tool to inject insight into the planning process, it is critically important to understand the limitations of such models.

## 2.1 The limitations of EEO models

Given the myriad simplifying assumptions required to make future energy-economy projections, a key concern is the ability of EEO models to deliver accurate results. We can in principle quantify the skill of a given model through validation exercises that compare model results with real world outcomes. Hodges and Dewar (1992) set forth four conditions that must be met in order for any model to be validatable:

1. It must be possible to observe and measure the situation being modeled.
2. The situation being modeled must exhibit a constancy of structure in time.
3. The situation being modeled must exhibit constancy across variations in conditions not specified in the model.
4. It must be possible to collect ample data with which to make predictive tests of the model.

Unfortunately, when applying the standards for model validation set forth by Hodges and Dewar (1992), EEO models fail Conditions 1, 2, and 3.[1] While it is possible to measure the situation being modeled, such as future energy prices, gross domestic product, or installed technology-specific capacity, the multi-decadal timeframes associated with most EEO model projections make observation practically impossible. For example, a projection model with a time horizon of 30 years requires us to wait 30 years to observe actual outcomes for validation purposes, which negates its purpose as a planning tool. Condition 2 requires that the "causal structure" of the system being modeled remain constant through time. Energy economies at different geographic scales also violate this condition. National priorities, technological change, and resource availability can result in structural economic shifts that are not captured by EEO models. As noted by Weyant (2009), a model that generated accurate predictions in the past may not work well in the future given underlying structural changes. Condition 3 requires that the system being modeled must not vary according to conditions that are not modeled. Since EEO models represent a highly simplified reality, there are many unmodeled conditions that can affect outcomes of interest, such as geopolitical events, major technological breakthroughs, environmental surprises, and shifting political moods over time. Since EEO models fail 3 of the 4 validation conditions, we conclude that there is no way to benchmark the performance of EEO models through rigorous validation exercises. And when retrospective analysis is performed, the general conclusion is that models have little skill in predicting outcomes of interest. (Craig et al., 2002; Morgan and Keith, 2008)

Given the inability to validate EEO model results against real world outcomes, model results are limited to the modeled world. Hodges and Dewar (1992) suggest that unvalidated models do not yield insight into the actual situation, only the assertions embodied in the model. Similarly, Morrison and Morgan (1999) assert that when building a model, modelers create "a representative structure" of a real world system. Consequently, model outputs pertain to the simplified modeled world rather than the real world. Models suggest truth rather than reveal it (Hodges and Dewar, 1992). As a result, while inter-model comparison is useful (e.g., Energy Modeling

---

[1] Craig et al. (2002) also apply Hodges and Dewar (1992) validation criteria to energy models used for projections. They conclude that conditions 2 and 3 are violated.

Forum, Innovation Modelling Comparison Project), it remains a weak form of validation because the model results are benchmarked against one another rather than real world outcomes.

*2.2  The importance of discoverable models and repeatable analysis*

While EEO model results are not self-evidently true and cannot easily be compared to real world outcomes through robust validation exercises, they still serve a vital role by testing and stretching our intuition regarding the function and response of energy systems and economies to policy and other external stimuli. Models serve an important cognitive function by allowing analysts to explore the decision space and observe the interactions among different components, which gives rise to model-based reasoning (Frigg and Hartmann, 2009).

A deeper question; however, pertains to exactly how analysts derive insight from models. Although focused on physics-based models, Hughes (1997) suggests that model-based learning takes place in three stages: denotation, demonstration, and interpretation. Likewise, Morrison and Morgan (1999) assert that much of the learning with models takes place through construction and manipulation. Unfortunately, the benefit of learning from model construction and application is hindered by the lack of openness characteristic of many existing EEO models, as detailed in Section 3.

In addition, the opaqueness of many EEO models prevents verification of model results by external parties. In the current context, verification represents the process of review that ensures that the data are error-free, assumptions are discoverable, and the mathematical model operating on the data produces the intended result. The inability to verify another's model results gives rise to three concerns:

1. Hidden flaws or bugs in the source code or data
2. Subjective or value-based assumptions driving the results
3. The effect of highly sensitive parameters obscured or absent in the published analysis

Public access to model source code and data would allay the concerns above by allowing external users to verify that the model works as intended, reproduce specific results, and test the robustness of model results through sensitivity analysis. While modeling teams ostensibly undergo an internal verification process, the inaccessibility of model source code and data prevents the same verification by external parties[2].

Furthermore, only by running the model can users understand the model world and how carefully crafted model structure, assumptions, and data affect model results. Allowing external parties to run a model under different assumptions also provides a check that the conclusions drawn from a published model-based analysis are indeed robust. Under the prevailing approach; however, the larger policy and modeling communities take a passive approach and rely almost exclusively on modelers to deliver policy-relevant observations. While it is unrealistic to assume that rigorous testing of model performance would be widespread with publicly available models, a small external group of expert users could enhance model credibility through verification exercises.

The practical necessity of verification exercises—repeatable experiments—is a core tenant of modern science (Henry, 1997). While some may argue that EEO models are not tools of traditional science and therefore exempt from scientific standards, we argue that the utility of a repeatability criterion is common to all forms of reason-based inquiry. The need for repeatable experiments at the dawn of the Scientific Revolution parallels the same need today with regard to EEO models. During the seventeenth century, the development of repeatable experiments laid the foundation for reason-based scientific inquiry by allowing independent verification of non-intuitive empirical results (Henry, 1997). During that period, the mathematical sciences played a substantive role in the development of the experimental method (Henry, 1997; Bennet, 1986). Unlike Aristotelian natural philosophy, which requires a premise based on self-evident experience, mathematics does not start from obvious premises, but rather requires acceptance of particular postulates (Henry, 1997). In addition, mathematics applied to real word phenomena can easily produce counter-intuitive results. Galileo Galilei (1564–1642), a pivotal figure in the scientific revolution of the seventeenth century (Machamer, 2010), provides a case in point. Galileo derived the well-known mathematical relationship that the distance traversed by an object in free fall is proportional to the square of the time spent in free fall. To

---

[2] It is worth noting that programming errors have an average occurrence rate of one to ten errors per thousand lines of code (Ince et al., 2012). Finding programming errors in closed models is likely to take longer, as less modelers have access to the model.

 

confirm the mathematical relationship and convince the skeptical, he repeated careful controlled experiments with falling bodies and inclined planes—creating "many memories of the same thing" —to demonstrate the explanatory power of his derived mathematical relationship (Dear, 1995; Drake, 1970). To imbue mathematical sciences with the same level of authority as natural philosophy, early experimentalists such as Blaise Pascal and Robert Boyle described their experiments in published literature (Henry, 1997). Although the style of exposition varied among them, the purpose of describing experiments was to convince a skeptical audience that a counter-intuitive result could be obtained by following a prescribed experimental procedure.

In order to be effective, an experiment had to be described in such a way that it was repeatable. It must be possible to check results time and again to prove their validity. As Henry (1997) states, "Experiments, like mathematics, are not self-evidently true. To be convinced of their truth, you either have to know what you are doing, or accept them on faith." Science is not based on faith, and repeatable experiments provide the means to confirm or falsify a theory, dataset, or model. Such experiments must be independently repeatable, or they carry no scientific authority.

Today, EEO models bear a resemblance to the mathematical sciences of the seventeenth century. EEO models consist of algebraic equations that represent a highly simplified and condensed version of reality. Like many mathematical inquiries, we must accept a set of preconditions that create a starting point for analysis that is not self-evident. In addition, EEO models often produce counter-intuitive insight. Rather than implicitly trusting published analysis, replication by external parties would provide an additional measure of confidence in the model results.

Several papers have noted the prevailing lack of transparency associated with the development and application of integrated assessment models (Ha-Duong, 2001; Rotmans, 1998; Schneider, 1997). Ha-Duong (2001) notes that journal referees are not required to verify the model results or audit the source code, but rather assess model-based analyses against existing literature and their own intuition. In addition, the reputation of the modeler(s) and the clarity of the associated narrative also play a role in assessment. However, if hidden simplifications, coding bugs, key sensitivities, or details drive specific results related to model formulation – given their often subjective nature – they would be impossible to identify by an external observer without a formal verification exercise. Schneider (1997) notes that integrated assessment models (IAMs) include value-laden assumptions that are often invisible to

   

the consumers of IAM results, who have widely varying analytic ability and command of model formulation. Such hidden assumptions confuse public debate and engender a sense of mistrust in black box models. Speaking more broadly about IA methods, Rotmans (1998) asserts that "the evidence for integrity is transparency." Repeatable analysis enabled by publicly accessible source code and data would increase transparency by allowing model interrogation and external verification of published results.

*2.3   A broad trend towards transparency in science*

The challenge of producing repeatable analysis exists for all scientific models, particularly those involving large datasets and/or complex model formulations. We note an increasing trend towards transparency as evidenced by the development of open source software across a range of scientific fields. For example, the Community Earth Science Model (CESM) in climate science (UCAR, 2012), an open source implementation of the Message Passing Interface (MPICH2) and the Portable Extensible Toolkit for Scientific Computation (PETSc) in computer science (MPICH2, 2012; PETSc, 2012), and the Basic Local Alignment Search Tool (BLAST) in bioinformatics (Altschul et al., 1990; BLAST, 2012) represent high visibility, open source efforts in their respective fields.

In addition, the importance of open databases has been recognized across scientific disciplines. For example, the GenBank sequence database is an open access, annotated collection of all publicly available nucleotide sequences and their protein translations (NCBI, 2012). Deep Sky is an astronomical image database of "unprecedented depth, temporal breadth, and sky coverage" gathered from the Near Earth Asteroid Tracking (NEAT) project (Deep Sky, 2012). Datasets from the Moderate-resolution Imaging Spectroradiometer (MODIS), a payload scientific instrument launched into Earth orbit by NASA, are publicly accessible (NASA, 2012). Recently, there have also been efforts to develop community science gateways (e.g., NERSC, 2012; XSEDE, 2012), which represent a community-developed set of tools, applications, and data that are integrated via a portal to meet the needs of a specific community.

There also appears to be a trend towards greater transparency in the academic literature. For example, the journal Science now requires the publication of computer code involved in the creation or analysis of data (Hanson et al., 2011). In addition, top economic journals, including American Economic Review (AEA, 2012),

Econometrica (TES, 2012), Journal of Political Economy (JPE, 2012), The Economic Journal (RES, 2012a), and The Econometric Journal (RES, 2012b) now have policies requiring the release of data, programs, and information required to replicate published results. Some research efforts go further by creating software elements that facilitate the re-creation of published results, such as the Executable Paper Challenge (Elsevier, 2012) and ReDoc (Schwab, 2000). Despite such progress, there are still many journals that have not implemented measures to ensure repeatability. For example, McCullough (2009) notes that replicable economic analysis in the peer-reviewed literature is the exception rather than the rule. Unfortunately, peer-reviewed archival journals that frequently publish EEO model-based results such as Energy Policy, Energy Economics, and Environmental Modeling & Assessment do not have policies requiring the publication of source code and data. Given the role of such models in public policy formulation, we find it problematic that these journals have not adopted the stricter accessibility standards of their peers.

## 3 An assessment of the current EEO model landscape

In order to assess the public accessibility of existing EEO models, we reviewed the annals of the International Energy Workshop (IEW) for the last decade, from 2001 to 2011 (IEW, 2012). The IEW is the premier international venue for energy modeling and analysis, and therefore represents an unbiased source for information on EEO models. We reviewed the conference programs and extracted model names from publicly accessible titles, abstracts, and papers. Because accessibility to conference abstract and papers is limited in some years, the review should be regarded as an unbiased sample rather than a comprehensive set. After completing our review, we found references to 43 EEO models. To make the analysis more tractable, consideration was limited to the models that were referenced in at least two different accepted IEW submissions over the last decade, which reduced the number of models from 43 to 12. Using the EEO model names identified in the IEW annals, we conducted a broad literature review to identify reports and journal articles describing the models. We tried to find the earliest descriptions and applications of the models, which are most likely to include a complete model formulation as well as source code and data. Table 1 provides information on the 12 models surveyed, including accessibility to model source code and data, software licenses, permission to redistribute, and commercial software dependencies.

 

Table 1 – Details on existing energy economy optimization models.

| Model | Model Type[a] | Public Source Code | Public Source Data | License | Redistribution[b] | Commercial Software | References |
|---|---|---|---|---|---|---|---|
| AIM | hybrid | Yes | Yes | No | Yes | Yes | Kainuma et al. (2003); AIM 2012 |
| DICE | CGE | Yes | Yes | No | Yes | Yes | Nordhaus (2011); Nordhaus (2008) |
| DNE21+ | TE/PE | No | No | No | NA | NA | Akimoto et al. (2004) |
| ERIS | TE | No | No | No | NA | Yes | Kypreos et al. (2000); Turton and Barrett (2004); |
| MARIA | CGE | No | No | No | NA | NA | Mori and Takahashi (1999); Mori (2000a); Mori (2000b) |
| MARKAL/TIMES[c] | TE/PE | No | No | Yes | No | Yes | ETSAP (2011); Loulou et a al (2004) |
| MERGE[d] | CGE | No | No | Yes | No | Yes | MERGE (2011); Manne and Richels (200-) |
| MESSAGE | TE/PE | No | No | No | NA | No | IIASA (2011); Messner and Strubegger (19) |
| OSeMOSYS | TE | Yes | Yes | Yes | Yes | No | OSeMOSYS (2011); Howells et al (2011) |
| SGM | CGE | No | No | Yes | No | No | JGCRI (2011); Brenkert (2005) |
| WEM | TE/PE | No | No | No | NA | Yes | IEA (2008); IEA (2011) |
| WITCH | hybrid | No | No | No | NA | Yes | Bosetti et al. (2006); FEEM (2012) |

[a] We distinguish two types of models: computable general equilibrium (CGE) and technology explicit, partial equilibrium (TE/PE).
[b] NA = Not applicable. Assigned when source and data are unavailable and no license is specified. Models with open source code and data but with no license are assumed to allow redistribution without any restrictions.
[c] MARKAL/TIMES: The datasets are developed by individual user groups, who make independent decisions about data availability.
[d] MERGE: The model is made available, but users must communicate with the modeling team to obtain access.

In order for a model qualify as having public source code or data in Table 1, it must be available for download through a publicly accessible web portal and without the need to communicate with the modeling team. Requiring an interested party to submit a formal request to obtain source code or data introduces a point of control that can either be used as an intentional filter to selectively deny access or as an unintentional barrier if the point-of-contact is non-responsive. We in no way suggest that these issues exist with the models reviewed below, but rather note that the possibility exists. As a result, we adopt a conservative metric when qualifying model source code and data as publicly accessible.

According to the accessibility standard described above, only AIM, DICE and OSeMOSYS provide direct public access to both model source code and data. Of those three, only OSeMOSYS has adopted a license that outlines how model source code and data may be redistributed[3]: the existing Apache version 2.0 license, which allows users the freedom to redistribute the model with minimal conditions (Apache, 2004). Neither AIM nor DICE specify a license, which implies that those models may be utilized and redistributed without any restrictions imposed by the original developers.

Looking across the 12 models presented in Table 1, it is clear that there are no community-wide standards regarding EEO model accessibility. Three-fourths of the

---

[3] Full disclosure: The first author is a member of the OSeMOSYS steering committee.

 

sampled models do not have publicly accessible source code and data. Many more EEO models not listed in Table 1 have limited or no public access. For example, within the U.S., both the Department of Energy and Environmental Protection Agency – government organizations with large modeling efforts – have not made model transparency a priority. Regarding the National Energy Modeling System (NEMS) used by the U.S. Energy Information Administration (EIA) to produce the *Annual Energy Outlook* (e.g., EIA, 2011a), the EIA notes that "[b]ecause of the complexity of NEMS, and the relatively high cost of the proprietary software, NEMS is not widely used outside of the Department of Energy" (EIA, 2009). While most NEMS source code and data is accessible after requesting a password (EIA, 2011b), little effort is made to foster third party verification. U.S. EPA's Office of Air and Radiation mainly uses consultants to develop and run closed source CGE models for use in economic analysis of climate policy. For example, MiniCAM is developed and run by the Joint Global Change Research Institute, the ADAGE model is developed and run by RTI International, and the IGEM model is developed and run by Dale W. Associates (EPA, 2011).

We posit several reasons why most EEO models and datasets remain closed source: protection of intellectual property, fear of misuse by uninformed end users, inability to control or limit model analyses, implicit commitment to provide support to users, unease about subjecting code and data to public scrutiny, and overhead associated with maintenance. We believe that, in the long run, the benefits of model transparency outweigh these legitimate concerns. Protection of intellectual property is a concern, but most of the intellectual value comes from the insight derived from model application, not the code and data. Model misuse is also a valid concern; however, publicly accessible code and data should enable the rapid identification of aberrant results. Regarding technical support, we agree with Barnes (2010) when he suggests that no one is entitled to free technical support, and constructive feedback can lead to model improvements while unhelpful feedback can be ignored. Finally, making a model publicly accessible can involve minimal overhead, as we outline in Section 5.

The existing open source models in Table 1 demonstrate that the concerns listed above are surmountable. DICE illustrates the benefit of a publicly accessible, transparent modeling effort: academic search engines return over 250 relevant references to journal articles, book chapters, and reports. (Figure 1). Given its greater complexity and regional focus, AIM has engendered less third party adoption, but

   12

has been used for high visibility policy analysis; for example, through the IPCC (Nakicenovic et al., 2000) and Stanford Energy Modeling Forum (EMF, 2012). Despite the recent development of OSeMOSYS (Howells, 2011), it has played a prominent role in the International Energy Workshop over the last couple of years with separate side events for interested modelers (IEW, 2012). We suggest that because the source code and data for these models are publicly accessible—and can therefore be carefully scrutinized—it engenders a sense of trust within the larger community of modelers that contributes to third party interest, scrutiny, and in many cases, adoption. In addition, publicly accessible models enable third parties to engage in model-based learning, as described in Section 2, without having to embark on a new model development effort.

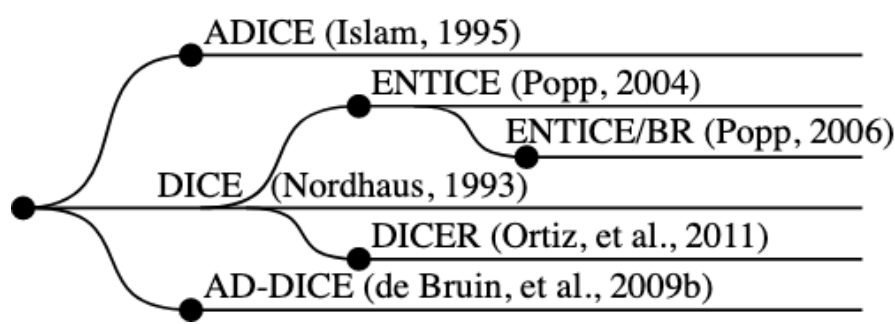

**Fig. 1.** Variants of the Nordhaus DICE model. A search for the "Dynamic Integrated model of Climate and the Economy" in Google Scholar returned 277 results on 7 April 2012. We reviewed all publication titles and abstracts to ensure relevance to the Nordhaus DICE model. Only DICE model variants with a unique name are included here, as they include modifications to the underlying DICE formulation rather than simply an application of the original formulation.

## 4  Keeping models in the public domain

As noted above, repeatable model exercises are only possible with access to model source code and data. Publicly accessible computer code is nothing new; there already exists a robust and vibrant community of open source software developers committed to software development in the public domain. The GNU Project, launched in 1984 to develop the GNU operating system, articulates a set of four freedoms that must be met in order to classify software as "free" (GNU, 2012a):

- The freedom to run the program, for any purpose (freedom 0).

 

- The freedom to study how the program works, and change it to make it do what you wish (freedom 1). Access to the source code is a precondition for this.
- The freedom to redistribute copies so you can help your neighbor (freedom 2).
- The freedom to distribute copies of your modified versions to others (freedom 3).

Note that the notion of *free* pertains to the lack of restrictions placed on the software, not the price. GNU defines *copyleft*—as opposed to *copyright*—as the general concept that protects these four freedoms (GNU, 2012a). However, implementation of copyleft requires a specific license, such as the GNU General Public License (GPL) or the GNU Lesser General Public License (LGPL). Perhaps the most visible example of free software is the Linux operating system, which is available through several different distributions (Negus, 2010). The Linux kernel, which represents the core operating system, is distributed under GPLv2 (Love, 2010). Several other open source licenses exist, including Apache, BSD, and Mozilla, which differ primarily over freedoms 2 and 3 regarding the conditions on redistribution.

The scientific projects mentioned in Section 2 (CESM, MPICH2, PETSc, BLAST, GenBank, DeepSky, and MODIS) are funded by the U.S. federal government and therefore do not place restrictions on the use and redistribution of available software tools and data. In contrast to EEO models, these datasets and software tools occupy a well-defined and focused space within the scientific community. By comparison, EEO models are often adapted to answer specific questions, require a large number of reasoned yet subjective judgments, and rely on input data that change frequently. Even ostensibly minor changes to input assumptions or model formulation can have unforeseen impacts on EEO model-based results. If source code and data associated with a particular analysis are not released, it is difficult for third parties to assess how changes made to the model affect the results, unless described by the modelers.

We assert that simply applying a copyleft license to an EEO model is insufficient. While freedoms 0 and 1 meet the stated objective of providing publicly accessible models, freedoms 2 and 3 fall short by relying on software redistribution to trigger the release of source code and data.

 14

To illustrate the consequences of unrestricted distribution, we performed a review of the DICE model, an open source EEO model drawn from Table 1. As Figure 1 suggests, DICE has been implemented several times with modifications and used to produce new model variants. While in all cases the authors made a good faith effort to describe changes to Nordhaus's original DICE model, none have made their source code and data publicly accessible. Even if DICE was subject to a copyleft license, the output of the modified model is simply a journal paper or report, which does not constitute software redistribution. As a result, the model user is not obligated to release the modified source code and dataset. In this case, the goal of repeatable model-based analysis cannot be sustained within the existing licensing framework.

Though not included in Table 1, the MIT Emissions Predictions and Policy Analysis (EPPA) model is noteworthy for its license agreement, which sets forth a set of special conditions (EPPA, 2011). Condition 6 within the license requires a user who publishes results based on a modification to the software must also "publish the source code of their modifications, in the same form as the Software Model is here released, and under the same license terms" (EPPA, 2011). Condition 7 states that redistribution of the "Software Model" must include the same license agreement (EPPA, 2011). The EPPA license is unique because it requires users to publish modifications to the model when model-based analysis is submitted for publication, and requires publication under the same EPPA license. Such a license fosters repeatable analyses by keeping model data and source code in the public domain.

There should be further discussion within the energy modeling community regarding the possible development of a new customized model license that promotes repeatable analysis through public accessibility. In order for such a customized software license to be effective, community vigilance is required and the threat of legal action for non-compliant modelers should be present. Research would be required to develop legally enforceable license language. Such a license agreement could be monitored and enforced through the peer review process. Even without a widely accepted license agreement, journals should be proactive by adopting a policy requiring the publication of source code, data, and information required to replicate a published model result.

## 5  Recommendations for EEO modelers

  

While Section 4 focused on the development and enforcement of open source standards from a community perspective, this section provides six best practice recommendations to modelers that can facilitate and sustain repeatable model analysis. Throughout this section, we suggest several common software development tools that can help facilitate community involvement with open source models.

*5.1 Recommendation 1: Make source code publicly accessible*

As noted in Section 2, all models represent a simplification of real world phenomena. In the case of EEO models, which project results over multiple decades, there are a large number of simplifying assumptions required to make the model formulation tractable. The resultant formulation is typically expressed as a set of equations in an algebraic modeling language (AML) such as GAMS (Rosenthal, 2008), AMPL (Fourer, 2002), or GEMPACK (CoPS, 2010). Algebraic languages allow for the formulation of an abstract model, which is distinct from the data on which it operates (Rosenthal, 2008). As a result, EEO model source code often represents a generic energy optimization framework, which operates on a specific dataset to produce results. While documentation, reports, and journal papers can help elucidate model formulation, only the source code represents the complete model implementation. This view is supported by Ince et al. (2012) who forcefully argue that even unambiguous descriptions of computer code are no guarantee of reproducibility.

EEO models with thousands of lines of code require auditing procedures that are transparent to both the modeling team and external observers. Software configuration management (SCM) is a software engineering practice to identify the configuration of a software system at distinct points in time for the purpose of systematically tracking and controlling changes to software (Bersoff, 1997). The goal is to ensure the integrity and traceability of changes during the entire software lifecycle.

Revision control, a major component of SCM, focuses on tracking the changes made to software, documents, or other digital information. It enables the versioned management of changes to a code base, analogous to "track changes" in a word processing document. A revision control system (RCS) logs all changes to a code base using incremental snapshots (known as "revisions") to prevent any loss of work. Moreover, best practices require the developer to briefly summarize the changes and provide semantic context for each snapshot. RCS also enables multiple developers to simultaneously work on a common software component and enhances productivity by automatically integrating changes made by different people. To serve that purpose, revision control systems support 'branching', a technique that enables simultaneous development of different branches emanating from the baseline repository. Once the prototype development is complete, the modifications from the branch can be integrated into the main repository.

Revision control systems also allow the creation of special software snapshots, called 'releases', which represent a well tested and clearly defined milestone in the software lifecycle. In the context of EEO models, such releases can be used to mark the software versions used to generate results for a specific publication or report. Revision control also facilitates internal auditing and, when coupled with web access, allows for easy external auditing of a code base (Read, 2003; Collins-Sussman, 2008; Loeliger, 2009). In our opinion, deployment of a revision control system via the web represents the most effective way to make the source code repository publicly accessible. This approach provides a transparent audit trail during model development and application that can quickly resolve potential controversies arising out of innocuous mistakes. While the code may be imperfect and the ability to provide support limited, public release enables other to engage in the research, which can ultimately lead to model improvements (Barnes, 2010).

In addition, as noted by DeCarolis (2010), there is a tendency to make models overly complex and data-intensive. EEO models should only be as complex as required to address the problem at hand. Rapid advances in computing power coupled with the growing abundance of data availability have engendered a philosophy of 'kitchen-sink' EEO models – the notion that a complex model should account for all relevant market dynamics. In contrast, modelers should strive for brevity and simplicity to promote transparency in model development. A revision control system, discussed above, allows modelers to develop alternative forms of the model with different features that suit a particular analysis and archive them as separate versions. The ability to archive different versions of the same model may help temper the tendency towards increasingly complex models over time as the same model is put to different uses. We emphasize; however, that revision control is simply a tool – the modeler must employ it judiciously with transparency as a design goal.

*5.2 Recommendation 2: Make model data publicly accessible*

Making source code publicly available does not make a model exercise repeatable unless the input data is also available. As noted in the previous section, EEO models programmed in algebraic modeling languages often maintain a distinction between the logic (i.e., model formulation) and data. As a result, algebraic models can operate on data files that vary by geographic region, technology detail, and parameter values. For example, the Energy Technology Systems Analysis Program (ETSAP) maintains the source code for two model generators, MARKAL and TIMES, but there are over 70 participating countries that have developed their own datasets for analysis (ETSAP, 2011).

While EEO model source code is likely to remain relatively stable over time, model data files are updated frequently as engineering parameters change, economic patterns shift, and the focus of analysis evolves. In order to generate reproducible model results, the complete set of model data should also be published in a web-accessible, free, and archived repository. Because input data and assumptions change so frequently, it is important to archive input data associated with each published analysis.

Most models store input data in text files, which can be easily integrated into existing revision control systems. MARKAL and TIMES require users to enter input data into a set of Excel workbooks, which are utilized to dynamically generate a relational database. At a minimum, a database (or the corresponding spreadsheets) could be archived in binary format. It may also be possible to extract the tables from each database version and archive them as text, which would enable line-by-line auditing of the data file by the revision control system.

There are ongoing efforts to make data publicly available for energy analysts and modelers. ETSAP has recently developed the Energy Technology Data Source (E-TechDS), which is comprised of technology briefs that contain cost and performance data relevant to energy models (ETSAP, 2012). Another open source data effort is megajoule.org, a wiki designed for energy analysts who wish to locate, contribute, and critique energy-relevant data (Henrion, 2011; Megajoule, 2012). The Global Trade Analysis Project (GTAP) maintains a global dataset that quantifies bilateral trade patterns as well as production, consumption, and intermediate use of services and commodities (GTAP, 2012). Versions of the GTAP database that are at least 2 versions old are publicly available for download (GTAP, 2012). While these efforts are a step in the right direction, we are not aware of any ongoing efforts to publicly

 

archive model- and analysis-specific data to enable repeatable, EEO model-based analysis.

*5.3 Recommendation 3: Make transparency a design goal*

Providing public access to model source code and data is a necessary but insufficient condition for model transparency. EEO models often include a complex algebraic formulation that operates on large datasets. While the availability of model source code and data allows external parties to reproduce earlier results, the user may have a very difficult time understanding the model without proper documentation.

An interested practitioner looks to documentation accompanying the model source code and data as the primary mechanism to aid model comprehension. The importance of documentation has already been recognized: all models listed in Table 1 include some form of documentation. Documentation comes in many forms; for example, a separate comprehensive document, comments embedded in the source code, and the code itself consisting of descriptive variable names that provide self-evident meaning. Well-articulated, accurate and current documentation should be a design requirement from the initial stages of model development. Where possible, it is also advisable to utilize automatic documentation generation tools (e.g., Doxygen, 2011; Sphinx, 2011) that process embedded documentation blocks in source code and are updated to reflect any source changes in order to maintain consistency between code and documentation. Additionally, modelers should consider providing different versions of documentation targeted for different purposes or users; for example, user and developer guides. For example, OSeMOSYS defines three levels of abstraction within the model framework: a plain English description, an algebraic formulation, and the model implementation in an AML (Howells et al., 2011).

*5.4 Recommendation 4: Utilize free software tools*

While we have emphasized free in the context of open source code and data, modelers should also consider the use of cost-free (gratis) software to minimize the barriers to entry. Open source programming environments for algebraic model formulation include GNU MathProg (GNU, 2012b), <Coliop|Coin> Mathematical Programming Language (CMPL) (COIN-OR, 2012a) and Python-based Optimization Modeling Objects (Pyomo) (Hart et al. 2012). All three of these software tools include linkages to free solvers. EEO models used to produce policy-relevant insight should maintain a baseline executable model that is (at a minimum)

© 2012. https://doi.org/10.1016/j.eneco.2012.07.004. This manuscript version is made available under the CC-BY-NC-ND 4.0 license http://creativecommons.org/licenses/by-nc-nd/4.0/    19

operable with free tools. In some cases, proprietary software may be preferable to solve more complex formulations or decrease solution time. For example, CPLEX often exhibits superior performance for linear and quadratic programming problems, but requires the purchase of an expensive (>$1000USD) commercial license (IBM, 2011). For those who cannot afford a CPLEX license, the model should be solvable with an open source solver.[4] Additionally, the cost of software impedes educational and research usage as it has a disproportionate impact on academics—particularly the ability to train the next generation of EEO modelers—and analysts in developing countries with minimal budgets. While the use of commercial software does not prevent repeatable analysis, the financial burden of software limits the number of individuals who can participate in the EEO modeling endeavor.

*5.5 Recommendation 5: Develop test systems for verification exercises*

To aid model development efforts, the EEO community should create a set of publicly available data files that represent test systems for verification exercises. Data files of varying complexity can be used to internally benchmark model results and performance as development proceeds, and can also be used for comparison to other models. For example, to debug the OSeMOSYS model, a test system called 'Utopia', packaged with MARKAL, was used to debug model formulation (Howells et al., 2011). While such an informal approach works, more beneficial would be a recognized suite of test systems. For example, in power engineering, the IEEE Subcommittee on the Application of Probability Methods, published the reliability test system (RTS) in 1979, which has subsequently been updated (Allan, 1986; Grigg et al., 1999). The IEEE RTS is recognized community-wide as a test system containing generation, transmission and load data that can be used for consistent evaluation of proposed reliability techniques (Allan, 1986). Such a suite of test systems within the international EEO community could serve several valuable

---

[4] There are many open source solvers.   Two examples are the GNU Linear Programming Kit (GLPK), which is licensed under the GPL and available at no cost (GNU, 2012b), and the suite of solvers available from COIN-OR (COIN-OR, 2012b).

 

purposes: debug changes to model formulation, assess computational performance, and provide a consistent evaluation mechanism for inter-model comparison.

*5.6 Recommendation 6: Work towards interoperability among models*

While collaborative modeling efforts such as the Energy Modeling Forum (2012) and Innovation Modelling Comparison Project (2011) are focused on drawing insight through inter-model comparison, participating modeling teams must expend significant effort to implement scenario assumptions within their respective input datasets. Such inter-model comparison is possible through the similarity of starting assumptions for analysis; however, it is difficult to determine whether divergent model results are due to differences in input data or model formulation. If models of the same class (e.g., technology explicit partial equilibrium or computable general equilibrium) could draw data directly from a common data repository, it would improve inter-model comparison by ensuring data consistency across models. In addition, if different models could share the same dataset, duplicative effort associated with building datasets could be reduced.

Various input formats already exist for sharing model data, including binary formats (e.g., HD5, NetCDF, XLS) and text formats (formatted ASCII, YAML, XML), but each is geared toward individual or archival data sets, rather than transparent comparison and global interoperability. In our view, a relational database management system (RDBMS) is the most appropriate tool to archive and query common model data. The advantage over other storage options is that data can be queried and mapped efficiently to a project's native input format without altering the underlying organization. The relationships are defined through a schema and enforced by the RDBMS (Imielinski, 1982).   Through the schema, an RDBMS provides a declarative layer of access to data: the user simply provides a descriptive definition of a model's data requirements in order to generate a model-specific dataset on demand. A key benefit of an RDBMS is that it works equally well for a dataset containing 10 or 10 billion data entries. Once developed and replicated at multiple institutions, a publicly accessible database could serve as a community resource for energy modelers.

# 6    Concluding thoughts

EEO models include complex energy system representations that operate on large datasets. The rise in computing power coupled with an increase in data availability has led to complicated models whose internal workings are often hidden from external observers. We assert that models should be fully discoverable by interested and knowledgeable third parties. Further, model-based analysis used to inform public policy should be independently reproducible by third party observers. Without the ability to repeat model analysis, it is impossible to fully understand a model formulation, expose hidden assumptions, or identify key model sensitivities. Currently, such analysis is assessed largely on the reputation of the modeling team, comparison to other analyses, and subjective interpretation based on the reader's experience and judgment.

In order to enable repeatable analysis with EEO models, modeling teams must publish both source code and data. While this paper provides several practical recommendations to facilitate open source models, we anticipate criticism of open models. A key concern is the misuse of EEO models either out of ignorance or purposefully by special interests. We believe that in the long run, an active international community of energy modelers using models with open source code and data will be self-correcting, as it is the case in the broader scientific community. Access to models can result in poor analysis, but access to the same models by others can ultimately identify poor assumptions driving model results. Such a position should not be controversial, as the requirement for repeatable experiments in modern science – dating back to the seventeenth century – has led to the identification of errors and the overall advancement of human understanding. An open process for model development and application would engender greater public trust by increasing transparency, produce a documented record of modeling insight, and catalyze development through the elimination of duplicative efforts related to data collection and model formulation.

**Acknowledgments**

This material is based upon work supported by the National Science Foundation under Grant No. CBET- 1055622. We also thank the anonymous reviewers for constructive feedback.This material is based upon work supported by the National Science Foundation under Grant No. CBET- 1055622. We also thank the anonymous reviewers for constructive feedback.